 \documentclass[preprint]{aastex}

\slugcomment{To appear in the Astrophysical Journal}

\shorttitle{Keplerian Disk around L1551 NE} 
\shortauthors{Takakuwa et al.}

\begin{document}


\title{A Keplerian Circumbinary Disk around the Protostellar System L1551 NE}
\author{Shigehisa Takakuwa\altaffilmark{1,2}, Masao Saito\altaffilmark{3}, Jeremy Lim\altaffilmark{4},
Kazuya Saigo\altaffilmark{3}, T. K. Sridharan\altaffilmark{5},\\
\& Nimesh A. Patel\altaffilmark{5}}
% Munetake Momose\altaffilmark{6}, \& Paul. T. P. Ho\altaffilmark{1}}

\altaffiltext{1}{Academia Sinica Institute of Astronomy and Astrophysics, P.O. Box 23-141, Taipei 10617, Taiwan}
\altaffiltext{2}{e-mail: takakuwa@asiaa.sinica.edu.tw}
\altaffiltext{3}{ALMA Project Office, National Astronomical Observatory of Japan, Osawa 2-21-1,
Mitaka, Tokyo 181-8588, Japan}
\altaffiltext{4}{Department of Physics, University of Hong Kong, Pokfulam Road, Hong Kong}
\altaffiltext{5}{Harvard-Smithsonian Center for Astrophysics, 60 Garden Street, MS78, Cambridge,
Massachusetts 02138, U.S.A.}
% \altaffiltext{6}{College of Science, Ibaraki University, Bunkyo 2-1-1, Mito, Ibaraki, 310-8512, Japan}

\begin{abstract}
% The transition from ``pseudo-disks'', which are commonly found around Class 0 protostars, to Keplerian disks,
% commonly found around pre-main-sequence (Class II) stars, is poorly understood, as are the
% properties of Keplerian disks during the transitional phase.

We present SubMillimeter-Array observations
of a Keplerian disk around the Class I protobinary system L1551 NE in 335 GHz continuum emission and
submillimeter line emission in $^{13}$CO ($J$=3--2) and C$^{18}$O ($J$=3--2) at a resolution of $\sim$120 $\times$ 80 AU. The 335-GHz dust-continuum image shows a strong central peak closely coincident with the binary protostars and likely corresponding to circumstellar disks, surrounded by a $\sim$600 $\times$ 300 AU feature elongated approximately perpendicular to the [Fe II] jet from the southern protostellar component suggestive of a circumbinary disk. The
$^{13}$CO and C$^{18}$O images confirm that the circumbinary continuum feature is indeed a rotating disk; furthermore, the C$^{18}$O channel maps can be well modeled by a geometrically-thin disk exhibiting Keplerian rotation. We estimate a mass for the circumbinary disk of $\sim$0.03-0.12 $M_{\odot}$, compared with an enclosed mass of $\sim$0.8 $M_{\odot}$ that is dominated by the protobinary system. Compared with several other Class I protostars known to exhibit Keplerian disks, L1551 NE has the lowest bolometric temperature ($\sim$91 K), highest envelope mass ($\sim$0.39 $M_{\odot}$), and the lowest ratio in stellar mass to envelope + disk + stellar mass ($\sim$0.65). L1551 NE may therefore be the youngest protostellar object so far found to exhibit a Keplerian disk. Our observations present firm evidence that Keplerian disks around binary protostellar systems, ``Keplerian circumbinary disks'', can exist. We speculate that tidal effects from binary companions could transport angular momenta toward the inner edge of the circumbinary disk and create the Keplerian circumbinary disk.
\end{abstract}

\section{Introduction}

% It is commonly accepted that protostar + disk systems are formed in the innermost part
% of protostellar envelopes \citep{and00,mye00}.
Previous interferometric observations of millimeter and submillimeter continuum emission
have shown compact ($\lesssim$600 AU) components
around both Class 0 and I protostars, embedded
in extended ($\gtrsim$2000 AU) envelopes \citep{loo00,ha03a,ha03b,eno09,eno11,jor09,mau10}.
Sub-arcsecond-resolution continuum observations that resolve the compact components reveal
flattened morphologies commonly attributed to circumstellar or possibly circumbinary disks
\citep{loo00,bri07,eno09,eno11}. The masses of the putative circumstellar disks
around Class 0 and I protostars are on the order of 0.01 - 1.0 $M_{\odot}$, much
higher than the typical masses of circumstellar disks around Class II (T-Tauri) stars
($\sim$0.005 $M_{\odot}$) \citep{and05,and07}. The sizes of the circumstellar disks ($\sim$600 AU)
around all these classes of stars, however, are comparable \citep{gui98,gui99,sch09}.

% Interferometric observations in molecular lines show that the circumstellar disks around Class 0 protostars
% are not rotationally supported ($i.e.$, do not exhibit Keplerian rotation). Instead, some show a strong infall
% component \citep{tak07,yen10,yen11}.

Previous interferometric observations
in molecular lines have shown an infalling motion and/or little rotating motion
in circumstellar disks around Class 0 protostars \citep{tak07,bri09,yen10,yen11}.
Such "pseudo-disks" have been predicted by theoretical simulations to form in the inner regions of envelopes
around protostars \citep{mel08,mel09,vor09,mac10,ma11a,ma11b}.
On the other hand, interferometric observations of T-Tauri stars in millimeter molecular lines
have revealed Keplerian disks around Class II stars \citep{gui98,gui99,pie03,sch09}.
These studies suggest that the transition from pseudo-disks to Keplerian disks occurs sometime
between the Class 0 and II stages.

Interferometric observations of molecular lines with, especially, the SubMillimeter Array (SMA),
which can (better) image higher transitions that trace gas at higher densities and/or temperatures,
have been increasingly successful at finding Keplerian disks around Class I protostars.
% Interferometric observations of molecular lines have been successful at finding Keplerian disks around Class I protostars.
% Brown and Chandler (1999) have reported Keplerian rotations around TMC-1 and TMC-1A with the dynamical masses
% of 0.2 $M_{\odot}$ - 0.4 $M_{\odot}$ and 0.35 $M_{\odot}$ - 0.7 $M_{\odot}$, from their OVRO observations of the
% $^{13}$CO (1--0) and C$^{18}$O (1--0) lines.
Brinch et al. (2007) have identified a (radius) $r$ $\sim$200 AU-scale Keplerian disk around L1489 IRS
with a central stellar mass of 1.35 $M_{\odot}$.
% from the HCO$^{+}$ (3--2) line and the detailed modeling of the envelope and the disk.
% The inferred disk radius and the disk mass are $\sim$200 AU
% and $\sim$4 $\times$ 10$^{-3}$ $M_{\odot}$, respectively.
Observations by Lommen et al. (2008) have also revealed a $r$ $\sim$100 AU-scale Keplerian disk
with a central stellar mass of 0.37 $M_{\odot}$ around IRS 63, and a $r$ $\sim$200 AU-scale Keplerian disk
with a central stellar mass  of 2.5 $M_{\odot}$ around Elias 29.
Observations by J$\o$rgensen et al. (2009) have shown a $r$ $\sim$140 AU-scale Keplerian disk around
the Class I protostar IRS 43 with a central stellar mass of 1.0 $M_{\odot}$.
The masses of these Keplerian disks range from $\sim$0.004 to $\sim$0.06 $M_{\odot}$,
slightly larger than that around T-Tauri stars,
whereas their radii are comparable to those of the Keplerian disks observed around T Tauri stars.
% Observations therefore suggest that the formation of Keplerian disks, or the transition from
% pseudo-disks to Keplerian disks, occurs sometime during the Class I stage.

% On the other hand, as mentioned above only the four, most-evolved Class I sources among the PROSAC samples show
% evidence for the Keplerian rotation in the HCO$^{+}$ (3--2) emission, where the central
% protostars possess 70 - 98$\%$ of the total (star + disk + envelope) masses.
% In other words, the disks found around the other, younger protostars are ``pseudo-disks''.

% Theoretical simulations indicate that relatively massive, ``pseudo-disks'' (disklike feature without Keplerian rotation)
% can form around protostars at early evolutionary stages, and could explain the presence of massive disks around young protostars
% \citep{mel08,mel09,vor09,mac10,ma11a,ma11b}.

Current theoretical simulations have trouble transforming pseudo-disks to Keplerian disks around protostars.
Mellon \& Li (2008; 2009) have performed ideal magnetohydrodynamic (MHD) simulations
of self-similar, collapsing envelopes to investigate formation of Keplerian disks.
They found that, for typical magnetic field strengths in interstellar clouds,
magnetic braking is so efficient that centrifugally-supported disks cannot form.
Machida et al. (2010; 2011a; 2011b) have conducted 3-D MHD simulations of
a collapsing Bonnor-Ebert sphere, including non-ideal MHD effects and finite spherical radii
($R_{\rm c}$ = 6.2 $\times$ 10$^{3}$ AU for a mass of 1.0 $M_{\odot}$). 
% They found that
% self-gravitating Keplerian disks can form around protostars until about 90$\%$ of the host
% cloud mass accretes onto either the protostar or circumstellar disk, or dissipates through
% outflows ($i.e.$, until the end of the main accretion phase or the Class I phase).
Their models show that the Ohmic dissipation and the depletion of the
surrounding infalling envelope that anchors the magnetic field significantly reduce the magnetic braking
and proceed the disk formation.
The predicted masses of the disks, however, are $\sim$a few to $\sim$a few hundred times
higher than the masses of the central protostar, and hence the disks are subject to further fragmentation.
Thus further long-term simulations are required
to reproduce the observed Keplerian disks around the Class I protostars.
To provide observational insights to theoretical models, it is important to
identify a Keplerian disk just after the transition from the pseudo-disk, and to
compare its properties to those of previously-identified Keplerian disks and the theoretical models.

Millimeter continuum observations have also revealed evidence for disks around
binary protostars, ``circumbinary disks'' \citep{loo97,loo00,mor00,lim06,mau10,che12}.
It is intriguing to investigate whether Keplerian rotation is present in those circumbinary
disks as well as disks around single protostars.
Although the Class I protostars around which Keplerian disks have been found are thought to be single,
it is possible that many and perhaps all are binary (or multiple) given the high frequency of binarity around
pre-MS and MS solar-mass stars \citep{duq91,del99,mat00}, as has been suggested for L1489 IRS
\citep{hog00,bri07}.
It is therefore important to compare the properties of circumbinary disks around known closely-separated
binary protostellar systems with the presumed circumstellar disks of apparently single protostars.
% Our ultimate goal
% is to unveil the unified picture of the circumbinary and circumstellar disks and
% the mass accretion flow from the circumbinary disk to each circumstellar disk \citep{bat97,och05}.

% primarily around single protostars,
% and there is so far no clear identification of Keplerian disks
% around (confirmed) binary protostars \citep{jor09}.
% Since the formation mechanism of binary protostars should be different from that of
% the single protostars (e.g., Nakamura \& Li 2003, Machida et al. 2008), the formation
% pathway of Keplerian disks around single and binary protostars may be different.
% It is thus important to know whether such disks exist around protobinary systems,
% the disk properties, and to compare with those around single protostars to see if any differences.

In this paper, we report observations of the Class I protostar L1551 NE in the 335 GHz continuum emission,
along with the $^{13}$CO ($J$=3--2) and C$^{18}$O ($J$=3--2) lines, at sub-arcsecond resolution.
Compared to the abovementioned Class I sources associated with the Keplerian disks,
L1551 NE is likely the youngest Class I object,
since the bolometric temperature ($T_{bol}$ = 91 K) \citep{fro05}
is more than a factor 3 lower and the amount of
the surrounding envelope around L1551 NE is
$\sim$one order of magnitude higher than that around the other protostars \citep{yok03,mor06,tak11}.
L1551 NE contains two 3.6-cm radio continuum sources with a projected
separation of $\sim$70 AU at a position angle of $\sim$300$\degr$: the south-eastern source
is referred to as ``Source A" and the north-western source ``Source B" \citep{rei02}.
Near-Infrared observations of L1551 NE have revealed that Source A drives collimated [Fe II]
jets along the north-east to south-west direction at a position angle of $\sim$60$\degr$
\citep{rei00,rei02,hay09}, and Source B is located at the origin
of an extended ($\sim$2000 AU) NIR reflection nebula \citep{rei00,rei02,hay09}.
There is little doubt that L1551 NE is a binary protostellar system.
Our observations of this nearby ($d$ = 140 pc; Elias 1978) object reveal for the first time a
Keplerian circumbinary disk.
In the following,
we shall describe observations ($\S$2), continuum and molecular-line results ($\S$3),
and the analysis to identify a Keplerian disk ($\S$4). In the last section ($\S$5), we will compare
the properties of the Keplerian disk around L1551 NE to those around the other protostars
and discuss the evolution and the formation mechanism of circumstellar / Keplerian circumbinary disks.

% These results suggest that Source A drives a collimated collimated jet while Source B
% a less-collimated, wide-angle wind, and that Source A is in the earlier evolutionary stage than Source B
% \citep{arc06}.
% We interpret these results as dierent evolutionary stages between Source A and Source B.
% Source A is in the earlier evolutionary stage, still strongly accreting from the circumstellar material,
% and drives a collimated [Fe II] (and possibly molecular) jet, whereas Source B is in the later
% evolutionary stage, less accreting, and drives a less collimated, fan-shaped molecular outflow.
%
% These results suggest that L1551 NE is a protostellar binary system.

\section{SMA Observations}

We observed L1551 NE with the extended configuration of the SubMillimeter Array (SMA)
\footnote{The SMA is
a joint project between the Smithsonian Astrophysical Observatory and the Academia Sinica Institute of Astronomy
and Astrophysics and is funded by the Smithsonian Institution and the
Academia Sinica.}
on 2010 January 15 and 18.
The $^{13}$CO ($J$=3--2; 330.587965 GHz) and C$^{18}$O ($J$=3--2; 329.3305453 GHz) lines, and
the continuum emission at both the upper and lower sidebands, were measured simultaneously.
The spectral windows (``chunks'') of the SMA correlator assigned to the $^{13}$CO and C$^{18}$O lines had
a resolution of 406.25 kHz, corresponding to a velocity resolution of $\sim$0.37 km s$^{-1}$.
In each sideband there are totally 48 chunks, providing a total bandwidth of 4 GHz.
All the chunks at both sidebands except for those assigned for the $^{13}$CO and C$^{18}$O lines
were combined to make a single continuum channel. The frequency coverages of the continuum observation
were 327.16 - 329.24 GHz ($\lambda$ = 0.911 - 0.916 mm), 329.34 - 330.47 GHz (0.907 - 0.910 mm),
330.57 - 331.13 GHz (0.905 - 0.907 mm), and 339.16 - 343.13 GHz (0.874 - 0.884 mm).
The total bandwidth of the continuum observation was 7.74 GHz. In the following, we refer to the continuum emission taken with the present
SMA observations as ``335 GHz continuum emission'', since the mean frequency of the combined continuum data is
$\sim$335.4 GHz. Details of the SMA are described by Ho et al. (2004).
Table 1 summarizes the observational parameters.
The minimum projected baseline length was $\sim$29 $k\lambda$ at the C$^{18}$O frequency,
and for a Gaussian emission distribution with a FWHM of $\sim$7$\arcsec$
($\sim$1000 AU), the peak flux recovered is $\sim$10$\%$ of the
peak flux of the Gaussian \citep{wil94}.
The estimated uncertainty in the absolute flux calibration is $\sim$30$\%$.
The raw visibility data were calibrated and flagged with MIR, which is an IDL-based data reduction
package \citep{sco93}. The calibrated visibility data were Fourier-transformed and CLEANed
with MIRIAD to produce the images \citep{sau95}.

\placetable{tbl-1}

\section{Results}
\subsection{335 GHz Continuum Emission}

\begin{figure}
\epsscale{1.0}
\plotone{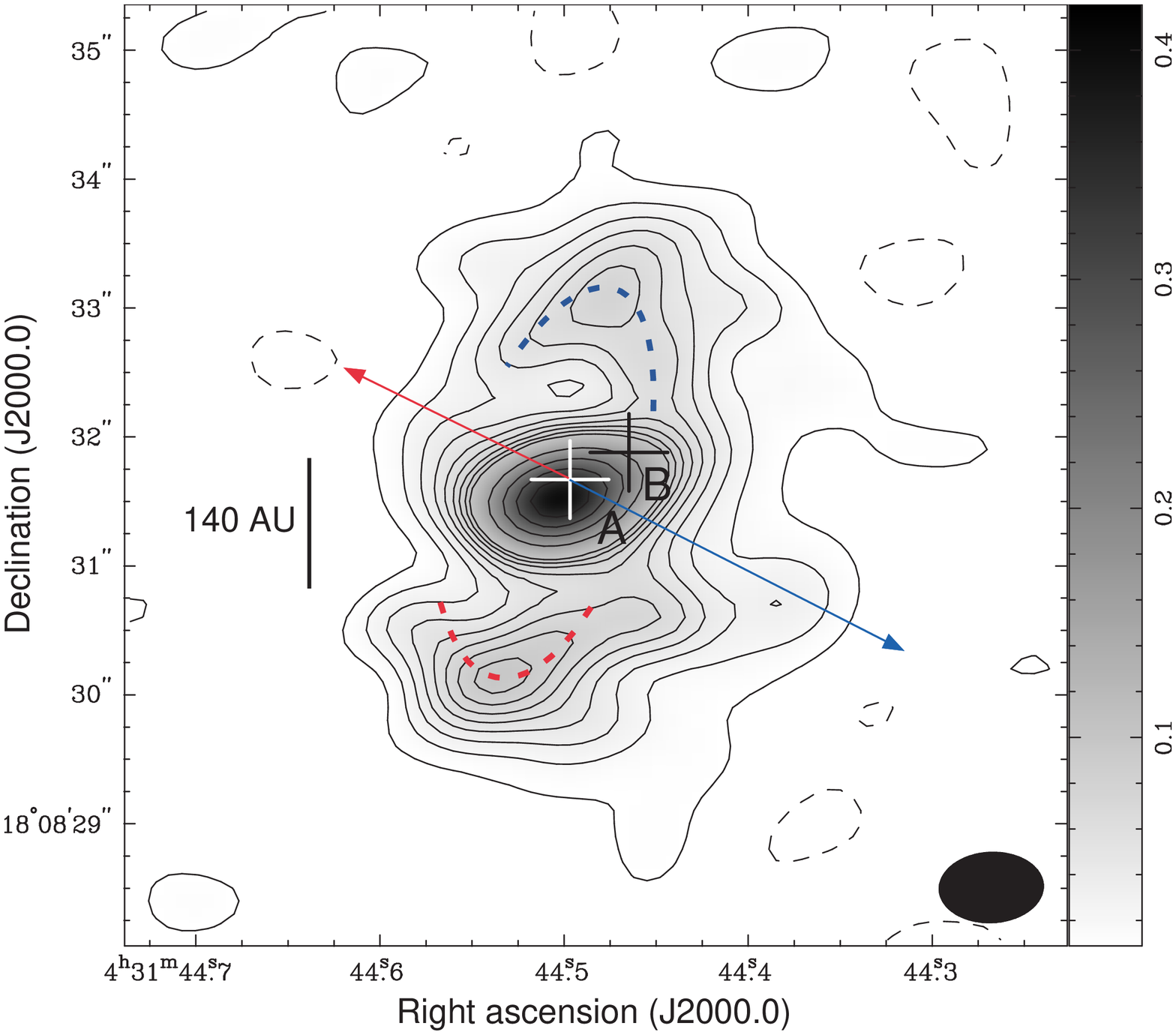}
\caption{335-GHz continuum image of L1551 NE observed with the SMA.
Contour levels are from 3$\sigma$ in steps of 4$\sigma$ until 35$\sigma$, and then in steps of
20$\sigma$ (1$\sigma$ = 3 mJy beam$^{-1}$). The highest contour level is 115$\sigma$.
Crosses indicate the positions of Source A and Source B
as labeled. A filled ellipse at the bottom-right corner shows the synthesized beam
(0$\farcs$80 $\times$ 0$\farcs$54; P.A. = -87$\degr$). Blue and red dashed curves delineate
the tilted $U$-shaped features, and blue and red arrows show the direction of the blueshifted
and redshifted [Fe II] jets driven by Source A.
\label{cont}}
\end{figure}

Figure \ref{cont} shows the 335 GHz continuum image of L1551 NE at an angular resolution of
0$\farcs$80 $\times$ 0$\farcs$54 (P.A. = -87$\degr$), corresponding to a spatial resolution of 110 $\times$ 80 AU.
The submillimeter continuum emission exhibits three distinct peaks.
The bright central peak is located $\sim$30 AU SE from Source A, and
exhibits a marginal extension towards Source B
(note that the central peak is elongated in a direction that is different to that of the synthesized beam).
This central component appears to be closely coincident with the protobinary.
From the 2-dimensional Gaussian fitting,
the total flux density of this component is measured to be $\sim$0.73 Jy and the deconvolved size
0$\farcs$90 $\times$ 0$\farcs$49 (P.A. = -72$\degr$).
The other two peaks, one to the north and the other to the south,
are located almost symmetrically with respect to the central component.
These peaks exhibit tilted $U$-shaped morphologies (see dashed curves in Figure \ref{cont}),
which trace a (diameter) $D$ $\sim$600 $\times$ 300 AU-scale inclined ring-like feature.
% approximately centered on the position of Source A.
The position angle of this feature ($\sim$160$\degr$) is approximately perpendicular to the axis of the [Fe II] jets
driven by Source A (P.A. = 64$\degr$; see blue and red arrows in Figure \ref{cont}).
The total flux density integrated over the entire emission region including these
three components is $\sim$1.65 Jy.
A similar three-peak feature, albeit at a lower S/N ratio, is also seen in the
1.3-mm continuum image of L1551 NE taken with the OVRO \citep{mor00}.
Without the benefit of observations in molecular lines, however, the interpretation made by Moriarty-Schieven et al. (2000)
for the continuum structure is very different from ours as described in Section 4.

% In $\S$5.2, we show that the central peak most likely compromises unequal contributions from the circumstellar disks
% associated with Sources A and B,
% and that the shell-like feature the rim-brightened edge of the circumbinary disk surrounding the protobinary system.

% One possible interpretation of these continuum features is that the central component represents
% the circumstellar disks around Source A and Source B, and that the shell-like feature traces the rim-brightened
% edge of the circumbinary disk surrounding the protobinary system. We will discuss this point
% in $\S$5.2 in more detail.
% Although the detailed emission distribution in the continuum emission
% is different from that of the C18O emission, the extent and the position angle of the continuum
% emission are consistent with those of the C18O emission. We consider that the submillimeter
% continuum emission also traces the circumbinary disk around the protobinary.

Single-dish observations of L1551 NE in the 850 $\micron$ continuum show
an extended (= 9$\farcs$8 $\times$ 8$\farcs$8) envelope with a total flux density of 2.78 Jy \citep{mor06}.
To assess the contribution from the extended envelope to the more compact emission observed with the SMA,
we created a 9$\farcs$8 $\times$ 8$\farcs$8 Gaussian image with a total flux density of 2.78 Jy,
and sampled the emission with the $uv$-coverage of our SMA observations.
We found that the envelope is almost entirely resolved out: we recovered a flux density of only $\sim$7.6 mJy,
less than 0.5$\%$ of the total flux density of the more compact emission observed with the SMA.
Thus, the contribution from the extended envelope to the more compact structure observed with the SMA is likely
negligible.
Similarly, J$\o$rgensen et al. (2009) found that the contribution from
extended envelope emission measured at mm/sub-mm wavelengths in single-dish observations to that in SMA observations having
baselines $\geq$50 $k\lambda$ is at most 8$\%$ for various envelope models.
We estimated a total (gas + dust) mass ($\equiv$$M_{g+d}$) for the feature detected in the 335-GHz continuum as shown in Figure 1 of
\begin{equation}
M_{g+d}=\frac{S_{\nu}d^2}{\kappa_{\nu} B_{\nu}(T_d)},
\end{equation}
where $\nu$ is the frequency, $S_{\nu}$ the flux density,
$d$ the distance, $B_{\nu}(T_d)$ the Planck function,
and $T_{d}$ the dust temperature, and $\kappa_{\nu}$ the dust opacity per unit gas + dust mass
on the assumption that the gas-to-dust mass ratio is 100. The frequency dependence of $\kappa_{\nu}$
can be expressed as $\kappa_{\nu}$ = $\kappa_{\nu_{0}}$($\nu$/$\nu_{0}$)$^{\beta}$, where $\beta$ denotes
the dust-opacity index.
By modeling the spectral energy distribution (SED) of the envelope as measured in
single-dish observations at wavelengths from 12 $\micron$ to 2 mm,
Moriarty-Schieven et al. (1994) derived $\beta$ $\sim$1 and 
Barsony \& Chandler (1993) and Moriarty-Schieven et al. (1994) $T_{d}$ $\sim$42 K.
If we adopt $\kappa_{\nu_{0}}$ = $\kappa_{1200 GHz}$ = 0.1 cm$^2$ g$^{-1}$ \citep{bec90}, then
from Eq. (1) we find $M_{g+d}$ $\sim$0.047 $M_{\odot}$. This value is comparable with that
estimated by Moriarty-Schieven et al. (2000) from their OVRO 1.3-mm image ($\sim$0.060 $M_{\odot}$)
based on the same parameters but after correcting for their different adopted distance of $\sim$160 pc.
J$\o$rgensen et al. (2009) performed a survey of low-mass protostars with the SMA and adopted a dust opacity of
of Ossenkopf \& Henning (1994) for grains with thin ice mantles coagulated at a density of $n_{H_{2}}$ $\sim$10$^{6}$ cm$^{-3}$;
$i.e.$, $\kappa_{850 \micron}$ = 0.0182 cm$^{2}$ g$^{-1}$ and $\beta$ = 1.7.
To permit a more direct comparison with the results of J$\o$rgensen et al. (2009),
we adopt this dust opacity law
and $T_{d}$ = 42 K, and from eq.(1) we find $M_{g+d}$ $\sim$0.078 $M_{\odot}$.
J$\o$rgensen et al. (2009) referred to
simulations by Visser et al. (2009) that showed dust temperatures at $r$=200 AU ranging from $\sim$20 to 50 K.
This range in dust temperatures gives $M_{g+d}$ = 0.063 $M_{\odot}$ ($T_{d}$ = 50 K) - 0.21 $M_{\odot}$ ($T_{d}$ = 20 K)
with the same dust opacity law of Ossenkopf \& Henning (1994).
These estimates show that $M_{g+d}$ is probably within the range of $\sim$0.05 - 0.2 $M_{\odot}$.
The same methods and
the flux density of the central component derived from the 2-dimensional Gaussian fitting ($\sim$0.73 Jy)
give the mass of the central component of $\sim$0.02 - 0.09 $M_{\odot}$.
Subtraction of this mass from $M_{g+d}$ gives the mass of the rest of the emission components
of $\sim$0.03 - 0.12 $M_{\odot}$.

\subsection{$^{13}$CO (3--2) and C$^{18}$O (3--2)}
%
% continuum peak ~0.2" SE from Source A
% 13CO single peak ~0.3" NW from Source A
% C18O twin peaks ~0.6" NW & ~1.2" SE from Source A
% Continuum ~1.5" NW & ~1.5" SE from Source A
%
% 13CO beam = 0.949 x 0.656 arcsec (P.A. = -87.6 deg) ---> peak S/N > 26 sigma ---> 0.949 / 26 ~ 0.0365" ~5 AU
% C18O beam = 0.953 x 0.659 arcsec (P.A. = -87.6 deg) ---> peak S/N > 10 sigma ---> 0.0953" ~13.342 AU
% continuum beam = 0.804 x 0.543 arcsec (P.A. = -87.3 deg) --> peak S/N > 115 sigma ---> 0.00699" ~0.9788 AU;  31 sigma, 23 sigma

\begin{figure}
\epsscale{1.0}
\plotone{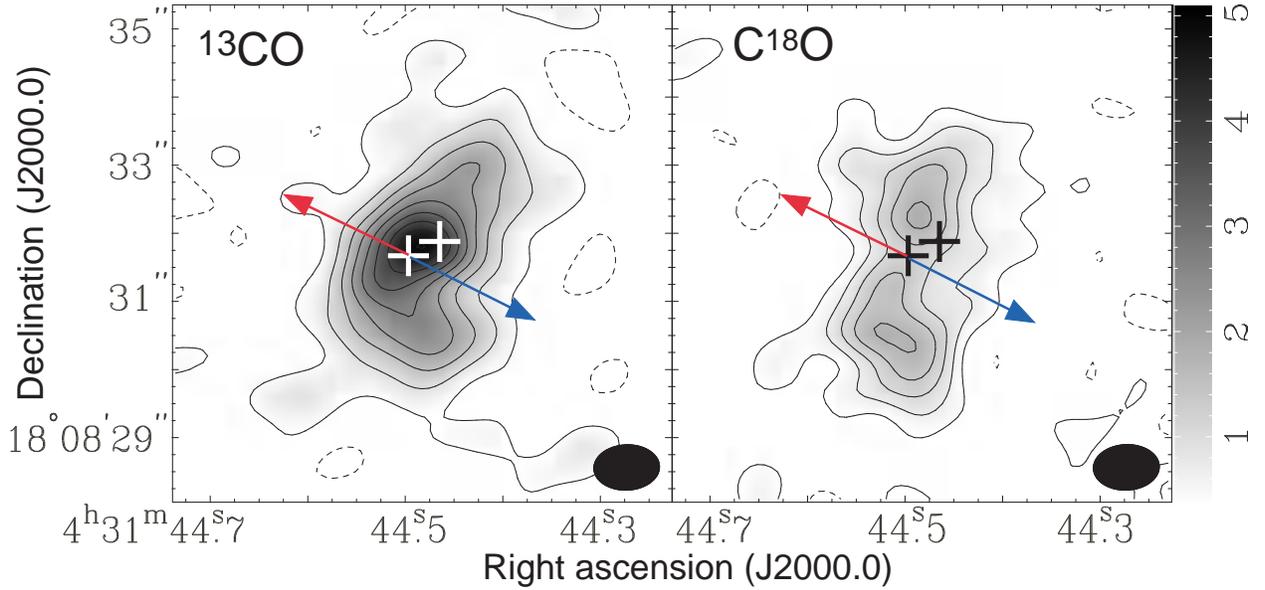}
\caption{Moment 0 maps of the $^{13}$CO (3--2) ($Left$) and C$^{18}$O (3--2) lines ($Right$)
in L1551 NE observed with the SMA. In the $^{13}$CO map,
contour levels are from 2$\sigma$ in steps of 3$\sigma$ (1$\sigma$ = 0.189 Jy beam$^{-1}$ km s$^{-1}$),
and the integrated-velocity range is 2.28 - 10.76 km s$^{-1}$.
In the C$^{18}$O map, contour levels are from 2$\sigma$ in steps of 2$\sigma$
(1$\sigma$ = 0.173 Jy beam$^{-1}$ km s$^{-1}$), and the integrated-velocity range is
3.72 - 9.63 km s$^{-1}$. Crosses indicate the positions of the protobinary,
and filled ellipses at the bottom-right corner the synthesized beam
(0\farcs95 $\times$ 0\farcs66; P.A. = -88$\degr$). Blue and red arrows show
the direction of the blueshifted and redshifted [Fe II] jets driven by Source A.
A righthand scale-bar shows the intensities of the gray-scale images in Jy beam$^{-1}$ km s$^{-1}$.
% Dashed lines show the major axis of the emission
% distributions, passing through the position of Source A.
\label{mom0}}
\end{figure}

Figure \ref{mom0} shows integrated intensity maps in $^{13}$CO (3--2) and C$^{18}$O (3--2).
In both $^{13}$CO and C$^{18}$O, the emission is elongated along the north-west to south-east direction.
This elongation is approximately perpendicular to the axis of the [Fe II] jets driven by Source A,
just like the ring-like structure seen in the continuum.
Furthermore, just like this structure, the emission in both lines
has outer dimensions of $\sim$600 $\times$ 300 AU.
% The extent ($\sim$600 AU $\times$ 300 AU) and the position angle ($\sim$160$\degr$) of these submillimeter
% molecular emission are similar to those of the 335 GHz continuum emission.
Although the overall sizes and position angles of the major axes of molecular and ring-like continuum structures are similar,
their internal structures are quite different. The three peaks seen in the continuum emission are not evident in
$^{13}$CO and C$^{18}$O. Instead, the $^{13}$CO emission exhibits a single peak between Source A and Source B
($\sim$40 AU NW from Source A), whereas the bright central peak in the continuum emission is located
$\sim$30 AU SE from Source A.
The positional accuracy of the $^{13}$CO emission peak can be approximately
estimated as $\sim$beam size / signal-to-noise ratio $\lesssim$5 AU,
and that of the continuum emission peak $\lesssim$1 AU. Thus, the positional offsets of the
$^{13}$CO emission peak from the continuum peak and the position of Source A are statistically significant.
By contrast, the C$^{18}$O emission does not show a central peak, but instead
shows two peaks at $\sim$80 AU NW and $\sim$170 AU SE from Source A, closer in than the two outer continuum peaks
which are located $\sim$210 AU NW and SE from Source A. The positional accuracy of these two C$^{18}$O emission peaks
is estimated to be $\lesssim$13 AU, and that of the two continuum peaks $\lesssim$5 AU. Therefore,
the offsets of the two C$^{18}$O emission peaks from the outer continuum peaks are also real.
The different emission distributions are likely due to the combination of
different amounts of envelope (and, in molecular lines, possibly also ambient) emission recovered, differences in optical depths,
and molecular abundance variations; these different effects are difficult to disentangle with the available data.

\begin{figure}
\epsscale{0.7}
\plotone{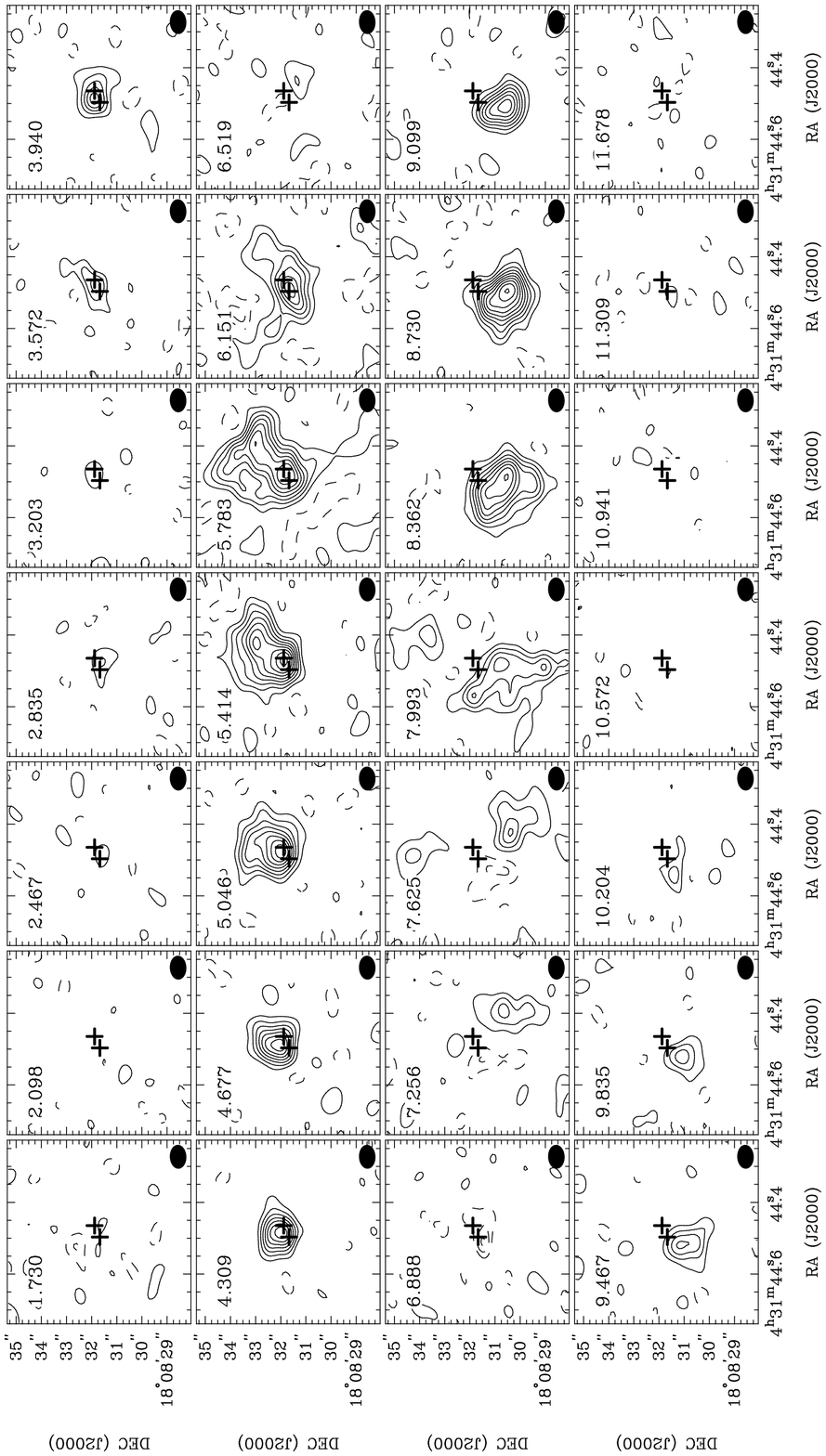}
\caption{Velocity channel maps of the $^{13}$CO (3--2) line in L1551 NE observed with the SMA.
Contour levels are from 2$\sigma$ in steps of 2$\sigma$ (1$\sigma$ = 0.107 Jy beam$^{-1}$).
Crosses indicate the positions of the protobinary, and a filled ellipse at the bottom-right corner in
each panel the synthesized beam
(0\farcs95 $\times$ 0\farcs66; P.A. = -88$\degr$). A number at the top-left corner in each panel denotes
the LSR velocity.
\label{ch13}}
\end{figure}

\begin{figure}
\epsscale{1.0}
\plotone{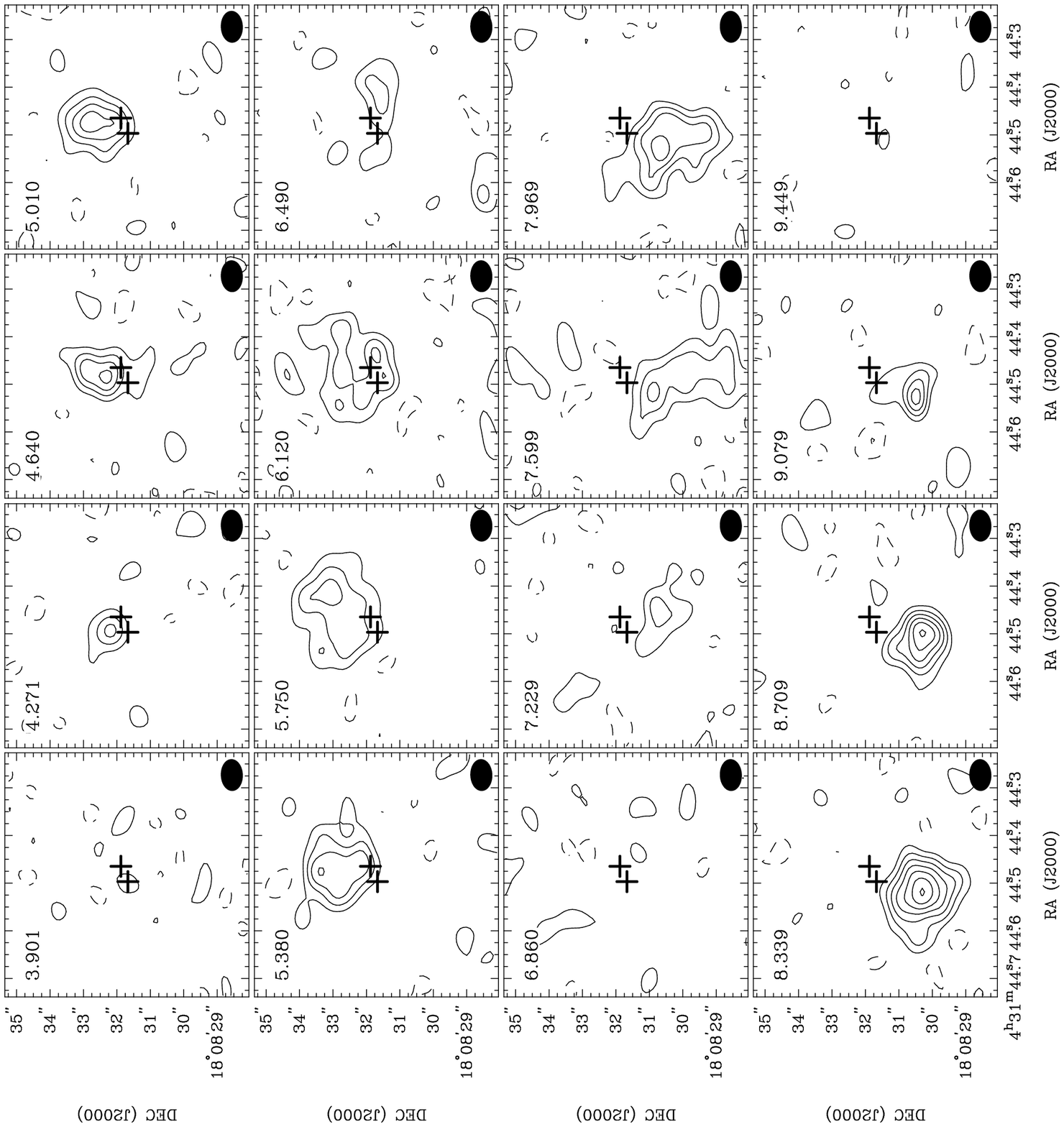}
\caption{Velocity channel maps of the C$^{18}$O (3--2) line in L1551 NE observed with the SMA.
Contour levels are from 2$\sigma$ in steps of 2$\sigma$ (1$\sigma$ = 0.117 Jy beam$^{-1}$).
Crosses indicate the positions of the protobinary, and a filled ellipse at the bottom-right corner in
each panel the synthesized beam
(0\farcs95 $\times$ 0\farcs66; P.A. = -88$\degr$). A number at the top-left corner in each panel denotes
the LSR velocity.
\label{ch18}}
\end{figure}

Figures \ref{ch13} and \ref{ch18} show velocity channel maps of the $^{13}$CO and
C$^{18}$O lines, respectively.
Around $V_{LSR}$ $\sim$7 km s$^{-1}$, both the $^{13}$CO and C$^{18}$O emission
are significantly suppressed, a likely effect of missing flux at the ambient velocity
of the molecular gas immediately around L1551 NE. Indeed,
single-dish CS (3--2, 5--4, 7--6) spectra towards L1551 NE show a central velocity
of $V_{LSR}$ $\sim$7 km s$^{-1}$ \citep{mo95b}.
A single-dish C$^{18}$O (3--2) spectrum taken towards L1551 NE with the CSO shows two distinct
components: one having a narrower velocity width ($\sim$0.68 km s$^{-1}$) with a central velocity
of $V_{LSR}$ $\sim$6.7 km s$^{-1}$, and the other having a broader velocity width ($\sim$2.2 km s$^{-1}$)
with a central velocity of $V_{LSR}$ $\sim$7.0 km s$^{-1}$ \citep{ful02}. The velocity range
and the central velocity of the broader component are similar to those of the
C$^{18}$O emission observed with the SMA.
We therefore adopt a systemic velocity for L1551 NE of $V_{LSR}$ = 7 km s$^{-1}$.
In the $^{13}$CO velocity channel maps, there are high-velocity
blueshifted (2.5 - 3.6 km s$^{-1}$) and redshifted (9.8 - 10.2 km s$^{-1}$)
components just displaced to the west and east of Source A, respectively.
These high-velocity $^{13}$CO components are not detected in the C$^{18}$O emission.
In the lower-blueshifted velocity ($\sim$3.9 - 5.8 km s$^{-1}$),
both the $^{13}$CO and C$^{18}$O emission are located predominantly to the north of the protobinary,
and in the lower-redshifted velocity ($\sim$7.6 km s$^{-1}$ - 9.5 km s$^{-1}$)
to the south of the protobinary.
Furthermore, the extents of the $^{13}$CO and C$^{18}$O emission appear to become
progressively larger as the velocity approaches the systemic velocity.
A comparison of the SMA C$^{18}$O (3--2) integrated spectrum with the CSO C$^{18}$O (3--2) spectrum
shows that from $\sim$3.9 to 5.4 km s$^{-1}$ and from $\sim$8.7 to 9.5 km s$^{-1}$ almost
100$\%$ of the total C$^{18}$O (3--2) flux is recovered in our SMA observations, although
the limited signal-to-noise ratio of the single-dish spectrum
at the line wings prevents us from accurately estimating the amount of the recovered flux.
Thus, the systematic increase in the spatial extent of the C$^{18}$O (3--2) emission over
these velocity ranges, where the detection level is more than 6$\sigma$ except for that at the
bluest and the reddest velocities, is probably real.
% In such a case, the extent and the velocity feature of the compact
% $^{13}$CO and the C$^{18}$O emission revealed with the present SMA observations may not be affected
% much by the contamination from the L1551 IRS 5 outflow, which is featureless in the present size scale.

\begin{figure}
\epsscale{1.0}
\plotone{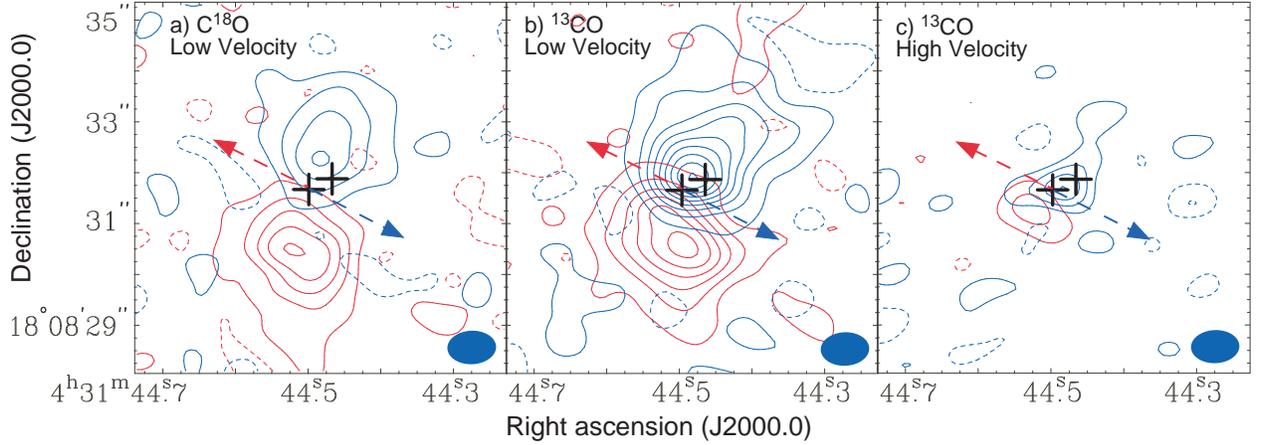}
\caption{a) Distributions of the low-velocity blueshifted (blue contours; 3.72 - 5.93 km s$^{-1}$) and
redshifted C$^{18}$O (3--2) emission (red contours; 7.41 - 9.63 km s$^{-1}$) in L1551 NE observed with the SMA.
Contour levels are from 0.212 Jy beam$^{-1}$ km s$^{-1}$ in steps of 0.424 Jy beam$^{-1}$ km s$^{-1}$
(1$\sigma$ = 0.106 Jy beam$^{-1}$ km s$^{-1}$). Crosses indicate the positions of the protobinary,
and a filled ellipse at the bottom-right corner the synthesized beam
(0\farcs95 $\times$ 0\farcs66; P.A. = -88$\degr$).
Blue and red arrows show the direction of the blueshifted and redshifted [Fe II] jets driven by Source A.
% A dashed line shows the major axis of the emission
% distributions, passing through the position of Source A.
b) Distributions of the low-velocity blueshifted (blue contours) and
redshifted $^{13}$CO (3--2) emission (red contours) in L1551 NE observed with the SMA.
The integrated velocity ranges, contour levels, and the symbols are the same as those in a),
while the 1$\sigma$ rms noise level is 0.097 Jy beam$^{-1}$ km s$^{-1}$.
c) Distributions of the high-velocity blueshifted (blue contours; 2.24 - 3.72 km s$^{-1}$) and
redshifted $^{13}$CO (3--2) emission (red contours; 9.63 - 10.37 km s$^{-1}$) in L1551 NE observed with the SMA.
Contour levels are from 0.158 Jy beam$^{-1}$ km s$^{-1}$ in steps of 0.158 Jy beam$^{-1}$ km s$^{-1}$,
where the 1$\sigma$ rms noise levels are 0.079 Jy beam$^{-1}$ km s$^{-1}$ and 0.056 Jy beam$^{-1}$ km s$^{-1}$
in the blueshifted and redshifted velocity ranges, respectively.
\label{br1318}}
\end{figure}

To better highlight the spatial-kinematic distributions of the different features
traced by the $^{13}$CO and C$^{18}$O emission,
we integrated these velocity channel maps over the four different velocity ranges
discussed above, that is,
high-blueshifted (2.5 km s$^{-1}$ - 3.6 km s$^{-1}$)
and redshifted velocities (9.8 km s$^{-1}$ - 10.2 km s$^{-1}$) where only $^{13}$CO emission is detected,
and low-blueshifted (3.9 km s$^{-1}$ to 5.8 km s$^{-1}$) and redshifted velocities
(7.6 km s$^{-1}$ to 9.5 km s$^{-1}$) where C$^{18}$O emission is also detected (Figure \ref{br1318}).
% The velocity range where bulk of the emission appears to be resolved out (6.67 - 7.41 km s$^{-1}$)
% is excluded in the integrated velocity ranges.
Overall, the low-velocity blueshifted and redshifted components in the $^{13}$CO and C$^{18}$O maps
are located to the north-west and south-east, respectively, of the L1551 NE system.
Furthermore, the axis connecting the blueshifted and redshifted emission peaks is approximately
perpendicular to the axis of the [Fe II] jets driven by Source A (see blue and red arrows in Figure \ref{br1318}).
Although the symmetric center of the blueshifted and redshifted emission appears to be closer to the position
of Source A, pinpointing the precise location of the symmetric center may not be straightforward as it seems.
% A single-dish C$^{18}$O (3--2) spectrum taken with CSO shows apparently two distinct
% Gaussian components; one with a narrower velocity width ($\sim$0.68 km s$^{-1}$) and a centroid velocity of
% $V_{LSR}$ $\sim$6.7 km s$^{-1}$ and the other with a broader velocity width ($\sim$2.2 km s$^{-1}$)
% and a centroid velocity of $V_{LSR}$ $\sim$7.0 km s$^{-1}$ \citep{ful02}.
%
% at 8 km s-1; SMA ~0.1 K; CSO ~0.25 K ---> 40 % recovered
% at 6 km s-1; SMA ~0.1 K; CSO ~0.43 K ---> 23 % recovered
% The narrower component in the CSO C$^{18}$O (3--2) spectrum, apparently
% originated from the ambient cloud component, is slightly blueshifted ($\sim$0.3 km s$^{-1}$)
% with respect to the another component which is likely to have the same origin of the SMA C$^{18}$O (3--2) emission.
A comparison with the CSO C$^{18}$O (3--2) spectrum indicates that
the amount of the missing flux at the blueshifted part of the SMA C$^{18}$O
emission is more than that at the redshifted part (for example, at $V_{sys}$ - 1.0 km s$^{-1}$
the amount of the missing flux is $\sim$80$\%$, while at $V_{sys}$ + 1.0 km s$^{-1}$ $\sim$60$\%$).
Furthermore, the location of L1551 NE overlaps with the spatially-extended
redshifted ($\sim$7 - $\sim$12 km s$^{-1}$) outflow driven by L1551 IRS 5 \citep{mor06}.
Thus the redshifted part of the $^{13}$CO emission may be more severely affected
by missing flux than the blueshifted part.
These ``asymmetric'' contributions from the extended components
may distort the ``balance'' between the blueshifted and redshifted emission distributions.
% and hence it is difficult to pinpoint the symmetric center of the blueshifted and redshifted emission.

In the high-velocity range, the compact ($\sim$100 AU) blueshifted and redshifted $^{13}$CO components
are located to the north-west and south-east of Source A, respectively.
% although the effect of the missing flux
% prevents us from identifying the dynamical center with respect to the binary positions.
The position angle of the axis
connecting the high-velocity blueshifted and redshifted $^{13}$CO components
is different from that of the low-velocity components; specifically, this axis is not orthogonal
to the axis of the [Fe II] jets driven by Source A.
These results imply that the origin of these high-velocity $^{13}$CO components is different from
that of the low-velocity components.
% We speculate that the circumstellar disks associated
% with Sources A and B may contribute to this high-velocity $^{13}$CO components, but otherwise do not consider
% the origin in the present paper.

\begin{figure}
\epsscale{1.0}
\plotone{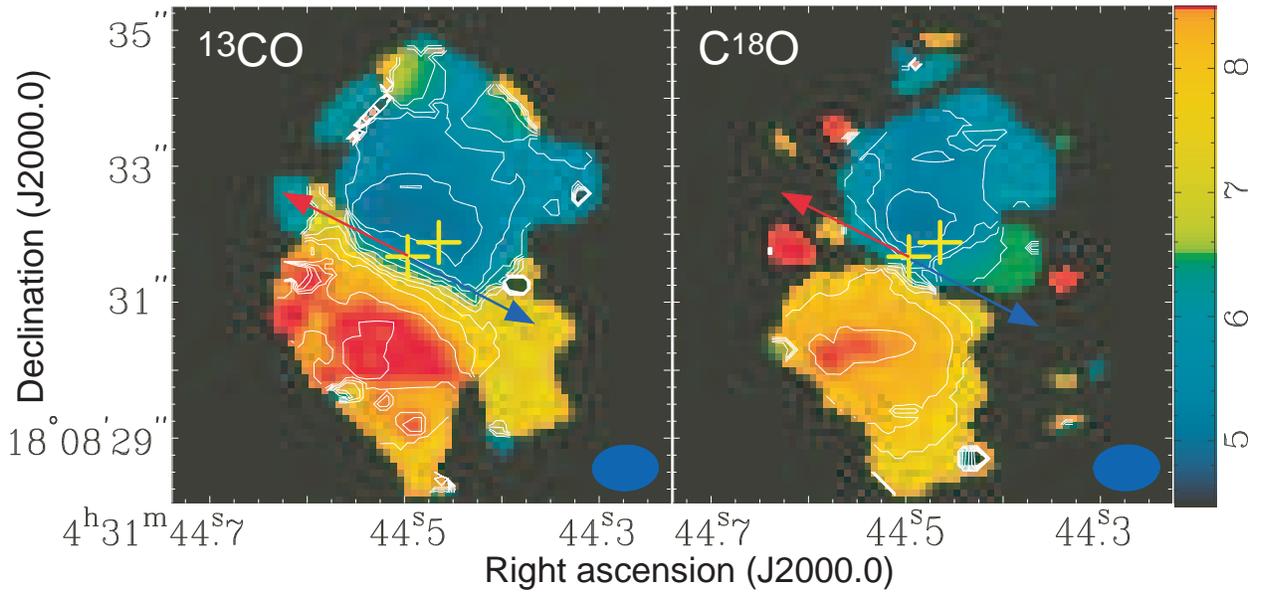}
\caption{Moment 1 maps of the $^{13}$CO (3--2) ($Left$) and C$^{18}$O (3--2) lines ($Right$)
in L1551 NE observed with the SMA. Contour levels are in steps of 0.4 km s$^{-1}$.
The bluest and reddest contour levels to the north-west and south-east of Source A in the $^{13}$CO map
are 5.2 km s$^{-1}$ and 8.8 km s$^{-1}$, respectively, and those in the C$^{18}$O map
5.2 km s$^{-1}$ and 8.4 km s$^{-1}$. Crosses indicate the positions of the protobinary,
and filled ellipses at the bottom-right corners the synthesized beam
(0\farcs95 $\times$ 0\farcs66; P.A. = -88$\degr$).
Blue and red arrows show
the direction of the blueshifted and redshifted [Fe II] jets driven by Source A.
\label{mom1}}
\end{figure}

Figure \ref{mom1} shows intensity-weighted mean-velocity maps of the low-velocity $^{13}$CO
and C$^{18}$O emission.
As mentioned above, the low-velocity blueshifted emission is distributed north-west
and redshifted emission south-east of the L1551 NE system.
We do not detect a velocity gradient along the minor axis of this structure indicative
of an inflow, or contamination by an outflow.
Thus, both the spatial and kinematic major axes of this feature are
approximately perpendicular to the axis of the associated [Fe II] jets (blue and red arrows in Figure \ref{mom1}).
% There is no clear velocity gradient along the minor axis,
% indicating no apparent contribution from a molecular outflow from L1551 NE.

\section{Analysis}

As shown in the previous section, the $^{13}$CO (3--2) and C$^{18}$O (3--2)
emission immediately surrounding L1551 NE exhibit a conspicuous velocity gradient along their
major axes perpendicular to the axis of the [Fe II] jets driven by Source A.
Such a velocity gradient could be produced by either a purely rotating or a rotating and
contracting ($i.e.$, infalling component) disk. To differentiate between these possibilities,
we tried model fittings of geometrically-thin Keplerian or infalling disks
to the observed $^{13}$CO and the C$^{18}$O channel maps.

Since we assume that the disk is geometrically thin, the centroid line-of-sight velocity at each position
in right ascension and declination with respect to the disk center ($\equiv$ $v_{LOS} (\alpha,\delta)$)
can be expressed as;
\begin{equation}
v_{LOS} (\alpha,\delta) = v_{sys} + v_{rot} (r) \cos(\Phi-\theta) + v_{rad} (r) \sin(\Phi-\theta),
\end{equation}
where
\begin{equation}
\Phi = \arctan(\frac{\alpha}{\delta}),
\end{equation}
\begin{equation}
r = \sqrt{(\frac{x}{\cos i})^2 + y^2},
\end{equation}
\begin{equation}
x = \alpha \cos(\theta) - \delta \sin(\theta),
\end{equation}
\begin{equation}
y = \alpha \sin(\theta) + \delta \cos(\theta).
\end{equation}
In the above expressions $v_{sys}$ is the systemic velocity, $\theta$ is the position angle of the disk major axis,
and $x$ and $y$ are coordinates of the disk along the minor and major axes, respectively.
$v_{rot} (r)$ denotes the rotational velocity of the disk as a function of the radius ($\equiv~r$).
In the case of Keplerian rotation, $v_{rot} (r)$ is expressed as;
\begin{equation}
v_{rot}(r)=\sin{i}\sqrt{\frac{GM_{\star}}{r}},
\end{equation}
where $i$ is the inclination angle of the disk from the plane of the sky, $G$ is the gravitational constant,
and $M_{\star}$ is the mass of the central star.
In the case of rotation with the conserved angular momentum, which is expected for an infalling disk \citep{oha97,yen11},
$v_{rot} (r)$ is expressed as;
\begin{equation}
v_{rot}(r)=\sin{i}\frac{j}{r},
\end{equation}
where $j$ denotes the specific angular momentum of the rotation.
In the case of the infalling disk, radial motion ($\equiv$ $v_{rad} (r)$) is also present
and can be expressed as;
\begin{equation}
v_{rad}(r)=\sin{i}\sqrt{\frac{2GM_{\star}}{r}}.
\end{equation}
The velocity channel maps of the model disk ($\equiv S_{model} (\alpha,\delta, v)$) can be expressed as;
\begin{equation}
S_{model} (\alpha,\delta, v) = (S_{mom0} (\alpha,\delta) / \sigma \sqrt{2\pi}) \times \exp(\frac{-(v-v_{LOS} (\alpha,\delta))^2}{2.0\sigma^2}),
\end{equation}
where $S_{mom0} (\alpha,\delta)$ denotes the moment 0 map of the model disk and $\sigma$ 
the internal velocity dispersion.

In the model fitting, $S_{mom0} (\alpha,\delta)$ was assumed to be
the same as the observed $^{13}$CO or C$^{18}$O moment 0 maps (Figure \ref{mom0}).
We adopted $\sigma$ to be 0.4 km s$^{-1}$, inferred approximately from the
$^{13}$CO and C$^{18}$O spectra.
These assumptions mean that we fit only the global, systematic
velocity structure in 3-dimensional space, and that we do not fit the 2-dimensional distributions of the molecular emission.
The center of the model disk was fixed to be the position of Source A, and $v_{sys}$ was set to be 7 km s$^{-1}$.
The $^{13}$CO and C$^{18}$O velocity channel maps around the systemic velocity
($V_{LSR}$ = 6.52 \& 6.89 km s$^{-1}$ for $^{13}$CO and $V_{LSR}$ = 6.86 \& 7.23 km s$^{-1}$ for C$^{18}$O)
were excluded from the fitting, since in these velocities the bulk of the emission is ``resolved out".
On these assumptions, we conducted minimum $\chi^2$-fittings of the model Keplerian disk
and the model infalling disk
to the observed $^{13}$CO and C$^{18}$O velocity channel maps,
with $M_{\star}$, $\theta$, $i$, and $j$ (in the case of the infalling disk) as fitting parameters;
\begin{equation}
\chi^2 = \sum_{\alpha,\delta,v} (\frac{S_{obs} (\alpha,\delta, v) -S_{model}^{M_{\star},\theta,i,j} (\alpha,\delta, v)}{\sigma_{rms}})^2 / \sum_{\alpha,\delta,v},
\end{equation}
where $S_{obs} (\alpha,\delta, v)$ denotes the observed velocity channel maps,
$S_{model}^{M_{\star},\theta,i,j} (\alpha,\delta, v)$ the model velocity channel maps with a given $M_{\star}$, $\theta$, $i$, and $j$,
and $\sigma_{rms}$ is the rms noise level of the observed velocity channel maps.

\begin{figure}
\epsscale{1.0}
\plotone{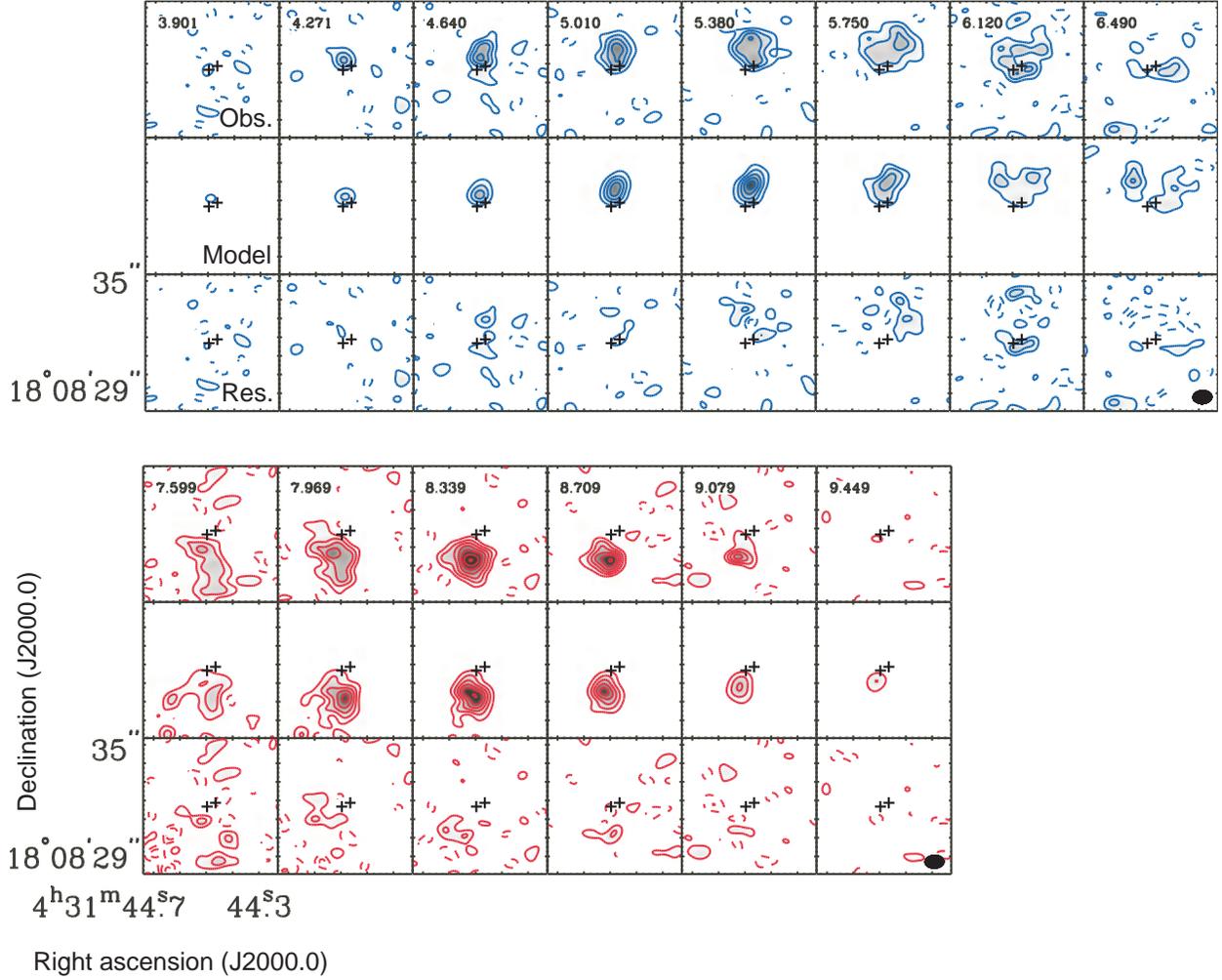}
% \plottwo{KepfitB.eps,KepfitR.eps}
\caption{Result of the $\chi^{2}$ fitting of the geometrically-thin Keplerian Disk to the
C$^{18}$O (3--2) velocity channel maps in L1551 NE at the blueshifted (blue contours) and redshifted (red contours)
velocities. Upper, middle, and lower panels show
the observed, model, and the residual velocity channel maps, respectively, where
the best fit parameters are $M_{*}$ = 0.8 M$_{\odot}$, $\theta$ = 167$^{\circ}$, and $i$ = -62$^{\circ}$.
Contour levels are from 2$\sigma$ in steps of 2$\sigma$ (1$\sigma$ = 0.117 Jy beam$^{-1}$).
Crosses show the position of the protobinary, and the filled ellipses at the bottom-right corners
the SMA synthesized beam (0\farcs95 $\times$ 0\farcs66; P.A. = -88$\degr$).
\label{kepb}}
\end{figure}

The middle and lower panels of Figure \ref{kepb} show the best-fit Keplerian-disk model and the
residual velocity channel maps of the C$^{18}$O emission, respectively, where the best-fit parameters are
$M_{\star}$ = 0.8$^{+0.6}_{-0.4}$ $M_{\odot}$, $\theta$ = 167$\degr$$^{+23\degr}_{-27\degr}$,
$i$ = -62$\degr$$^{+25\degr}_{-17\degr}$.
The residual velocity channel maps do not show any systematic features
although there are occasional 4$\sigma$ peaks; such random features with 4$\sigma$ peaks
are also seen in the Keplerian model to DM Tau by Guilloteau \& Dutrey (1998).
The best-fit residual rms is $\sim$0.144 Jy beam$^{-1}$, which is slightly higher
than but comparable to the noise level of the observed C$^{18}$O
velocity channel maps (1$\sigma$ $\sim$0.117 Jy beam$^{-1}$).
Our model fit of an infalling disk to the C$^{18}$O velocity channel maps, on the other hand,
cannot find the local $\chi^2$ minimum
within the range of $\theta$ from 109$\degr$ to 199$\degr$;
$i.e.$, the disk major axis is approximately perpendicular to the axis of the [Fe II] jets
and more than 45$\degr$ away from the jet axis. These results show that
a simple geometrically-thin Keplerian-Disk model with only three fitting parameters
reproduces the global velocity feature in the C$^{18}$O velocity channel maps.
The position angle of the Keplerian rotation is consistent
with the major axis of the C$^{18}$O total integrated intensity map, and is approximately
perpendicular to the axis of the [Fe II] jets driven by Source A.
The estimated inclination angle of the disk from the plane of the sky is also approximately
perpendicular to the inferred inclination angle of the [Fe II] jets of $i$ = 30$\degr$ - 45$\degr$ \citep{hay09}.
The position angle and the extent of the Keplerian disk in the C$^{18}$O emission
are similar to those of the ring-like structure in the 335 GHz continuum emission surrounding the binary protostars.
These results show that the circumbinary continuum feature is indeed a Keplerian disk surrounding the L1551 NE
binary system, or ``Keplerian circumbinary disk''.
Interestingly, the estimated position and inclination angles of the circumbinary disk
around L1551 NE are similar to those of the circumstellar disks \citep{lim06}
and the protostellar envelope \citep{mom98,tak04} around L1551 IRS 5, but their rotation
is in the opposite directions (in L1551 IRS 5 the northern part is redshifted).

By contrast, we could not find a satisfactory Keplerian fit (the peak residual is more than 10$\sigma$)
to the $^{13}$CO velocity channel maps. The same Keplerian-disk parameters do reproduce the overall
spatial-velocity distributions in the $^{13}$CO velocity channel maps, albeit with large residuals,
presumably because the $^{13}$CO emission is optically thick and likely affected
by the contamination from the extended envelope component and possibly the outflow from L1551 IRS 5.
Our simple geometrically-thin disk model,
which implicitly assumes optically-thin emission, cannot therefore be accurately applied for the $^{13}$CO emission.

\begin{figure}
\epsscale{1.0}
\plotone{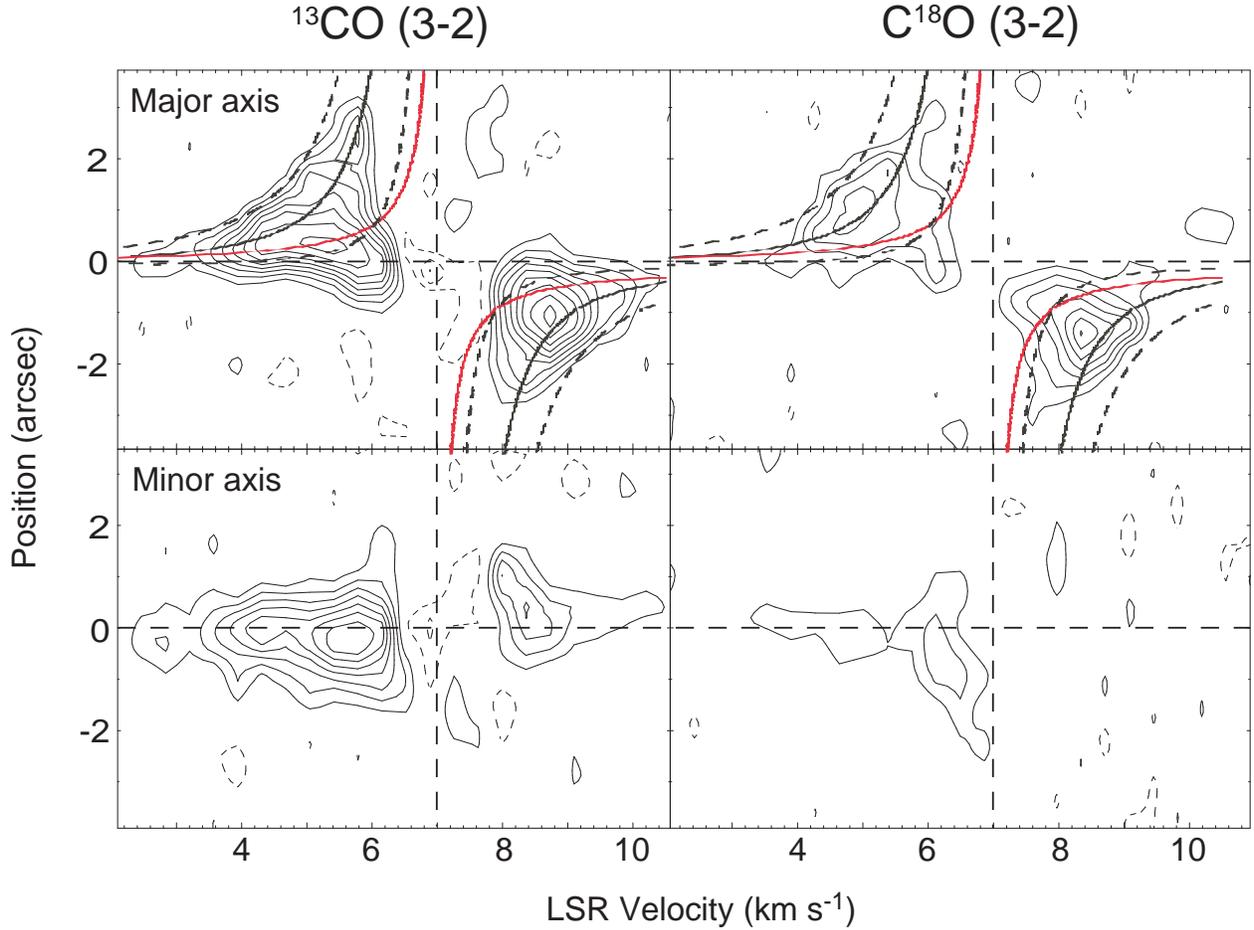}
\caption{Position - Velocity (P-V) diagrams of the $^{13}$CO (3--2) ($Left$) and C$^{18}$O (3--2) lines ($Right$)
along the major (P.A. = -13$\degr$; $Upper$) and minor axes (P.A. = 77$\degr$; $Lower$) of the best-fit model Keplerian disk,
passing through the position of Source A. Contour levels are from 2$\sigma$ in steps of 2$\sigma$
(1$\sigma$ = 0.107 Jy beam$^{-1}$ and 0.117 Jy beam$^{-1}$ for the $^{13}$CO and C$^{18}$O P-V diagrams,
respectively). Horizontal and vertical dashed lines denote the position of Source A and systemic velocity
of 7 km s$^{-1}$, respectively.
Solid and dashed black curves show the best-fit Keplerian rotation curve and the upper and lower ends
of the rotation curve derived from the error bars of the fitting parameters, while
red curves denote rotation with the conserved specific angular momentum
expected in infalling gas (see text).
\label{pv}}
\end{figure}

Figure \ref{pv} shows Position - Velocity (P-V) diagrams of the $^{13}$CO ($left$ $panels$)
and C$^{18}$O emission ($right$) along the major ($upper$) and minor axes ($lower$) of the model Keplerian disks,
passing through the location of Source A.
In the P-V diagrams along the major axis, the blueshifted and redshifted components are
distinctly separated and located to the north and south, respectively, of L1551 NE.
% while there is no apparent velocity gradient along the minor axis.
The emission at higher velocities is located closer to Source A than that at lower velocities,
with this trend more pronounced in the $^{13}$CO line.
The best-fit Keplerian-rotation curve is shown overlaid as solid black curves, and the upper and lower
ends of the rotational velocity derived from the error bars of the fitting parameters from
the $\chi^2$-fitting overlaid as dashed black curves.
% The possible range is calculated
% from the possible parameter ranges deduced from the $\chi^2$-fitting,
% and from the maximum and minimum values of $\cos(\theta)$$\sin(i)$$\sqrt{M_{\star}}$.
For comparison, a rotation curve with conserved angular momentum as expected for infalling gas
($j$ = 6$\times$10$^{-4}$ pc km s$^{-1}$; red curve) is also plotted.
Although both black and red curves trace the higher-velocity parts close to the center
($V_{LSR}$ $\sim$2.5 - 4.5 km s$^{-1}$ \& 9 - 10 km s$^{-1}$) equally well,
the Keplerian-rotation curve better traces the
lower-velocity parts ($\sim$5 - 6 km s$^{-1}$ \& 8 - 9 km s$^{-1}$).
We note that the effect of the missing flux may distort the true rotation curves.
% ?5.4 km s?1 and from ?8.7 km s?1
Keplerian-rotation curves ($v_{rot} \propto r^{-0.5}$) are, however, shallower than rotation curves
with conserved angular momenta ($v_{rot} \propto r^{-1}$), and thus the spatial extent
of the Keplerian disk at a given velocity is larger, as shown by the black and red curves.
Hence, it is possible that a genuine Keplerian rotation mimics a $r^{-1}$ rotation
due to the effects of missing flux, but it is unlikely that the effect of the missing flux
distorts a genuine $r^{-1}$ rotation to a Keplerian rotation.

% Along the minor axis, no clear velocity gradient is seen in the $^{13}$CO emission.

% We suggest that there is a Keplerian circumbinary disk around the protobinary system of L1551 NE.
% The central mass derived from the Keplerian fitting ($\sim$0.8 $M_{\odot}$) is likely
% the mass of the protobinary, since the mass of the circumbinary disk derived from the 335 GHz continuum emission
% ($\sim$0.047 $M_{\odot}$) is much smaller.
% Furthermore, it is likely that the inferred central mass is close to the mass of Source A and that mass of Source B is,
% by comparison, much smaller as the fixed rotation center at the position of Source A provides the decent fitting.
% be located close to Source A.

% Although the detailed emission distribution in the continuum emission
% is different from that of the C18O emission, the extent and the position angle of the continuum
% emission are consistent with those of the C18O emission. We consider that the submillimeter
% continuum emission also traces the circumbinary disk around the protobinary.
% The northern and southern continuum peaks with shell-like features may represent rim-brighted
% features of the Keplerian circumbinary disk, while the intense central component the circumstellar disk
% around each binary companion.

\section{Discussion}
\subsection{Evolution of Disks around Protostars}

% DM Tau Rout=850 AU
% GG Tau Binary Rout =260 AU
% TMC-1A ?
% L1551 IRS 5 j=8.1 x 10-4 pc km s-1
% It is probably difficult at this moment to say that binary has higher angular momentum....
% and to say that the large Keplerian radius, compared to Yen et al. single star radius, is intriguing.
%
% L1551 NE envelope mass ~0.39 M, smaller than the binary mass!! (Moriarty-Schieven et al. 2006)
% c.f. L1551 IRS 5 envelope mass ~1.01 M

Our SMA observations of L1551 NE in the $^{13}$CO and C$^{18}$O (3--2) lines
have revealed a Keplerian disk around the protobinary system. From the outer extent of the molecular-line
distributions, the outermost detectable radius of the Keplerian disk is estimated to be $\sim$300 AU.
The ring-like component seen in the 335 GHz continuum emission likely traces this Keplerian circumbinary
disk, while the central component may arise from circumstellar disks around the binary companions.
The central mass derived from the Keplerian fitting ($\sim$0.8 $M_{\odot}$) is likely
the mass of the protobinary, since the mass of the Keplerian circumbinary disk plus circumstellar disks derived from the 335 GHz continuum emission ($\sim$0.047 $M_{\odot}$) is much smaller.

Previous interferometric observations of Class 0 protostars have revealed
infalling motion in compact ($\lesssim$600 AU) circumstellar structures; $i.e.$, peudo-disks
\citep{tak07,bri09,yen10,yen11}. On the other hand, the large SMA
survey for protostellar sources (``PROSAC'' project) has found
Keplerian disks in the HCO$^{+}$ (3--2) line around four Class I protostars
(L1489-IRS, IRS 63, Elias 29, and IRS 43) out of their ten Class I samples
\citep{bri07,lom08,jor09},
one of which (L1489-IRS) was reported as a protobinary \citep{hog00}.
In the other six Class I sources the velocity fields traced by the HCO$^{+}$ (3--2) line are rather complicated, presumably due to the contamination from the outflows.
The reported Keplerian disks have radii $\sim$100 - 200 AU and masses $\leq$0.007 to 0.055 $M_{\odot}$,
and enclosed stellar masses 0.37 - 2.5 $M_{\odot}$.
These results imply that Keplerian disks can be present around Class I protostars
as well as T-Tauri stars (e.g., Guilloteau \& Dutrey 1998; Guilloteau et al. 1999).
The measured physical properties of these protostars and the Keplerian disks, including those of L1551 NE,
are summarize in Table \ref{tbl-2}.

Table \ref{tbl-2} shows that there are no clear differences of the disk radius and the enclosed stellar mass
between L1551 NE and the other sources. On the other hand, the bolometric
temperature of L1551 NE ($T_{bol}$ = 91 K) is a factor $\gtrsim$3 lower than that
of the other protostars, and the mass of the surrounding envelope around L1551 NE is
a factor of $\sim$4 higher than that of L1489-IRS and over an
order of magnitude higher than that around the other protostars.
Furthermore, the fraction of the central protostellar mass to the total (star + disk + envelope) mass
is the lowest in L1551 NE (= 65$\%$).
These results imply that
either L1551 NE will evolve into a much more massive star compared with
the other protostars listed in Table \ref{tbl-2}, or L1551 NE is the youngest Class I protostar yet found to have a Keplerian disk.
Since the L1551 region forms solar-type ($\lesssim$1 $M_{\odot}$) stars \citep{hay93,mom98,tak04,hay09},
the former case is unlikely.
% Only the four, most-evolved Class I sources among the PROSAC samples show
% evidence for Keplerian rotation, and 
% These results imply that
% during the Class I stage ``transition'' from pseudo-disks to Keplerian disks occurs,
Thus, we suggest that L1551 NE has experienced ``transition'' from an infalling,
pseudo-disk to a Keplerian disk most recently among the Class I sources
with the Keplerian disks.
The formation mechanism of Keplerian disks around protostars will be discussed in the next section.

\subsection{On the Formation of a Keplerian Circumbinary Disk}

Our SMA observations of L1551 NE provide firm evidence that
Keplerian circumbinary disks are present in addition to Keplerian circumstellar disks around Class I protostars.
On the theoretical side, however, it is still difficult to form Keplerian disks around either single or multiple protostars.
Mellon \& Li (2008; 2009) have conducted 2-D axisymmetric simulations of the collapse of
rotating, magnetized singular isothermal cores.
They found that, even in the case of weak initial magnetic fields (mass-to-flux ratio $\lambda$ $\lesssim$100),
magnetic fields efficiently carry angular momenta outwards, preventing the formation of a rotationally-supported disk.
They showed that a field strength in the dense core L1544 inferred observationally ($\lambda$ $\lesssim$10; Crutcher \& Troland 2000)
completely suppresses the formation of the centrifugally-supported disk, even if
the ambipolar diffusion is taken into account. They concluded that additional processes,
such as Ohmic dissipation, Hall effect, or the dispersion of the surrounding envelope that anchors the magnetic brake,
are required for a Keplerian disk to form.
Machida et al. (2010; 2011a; b) have performed 3-D magnetohydrodynamic
simulations of a Bonnor-Ebert sphere with a mass of 1 $M_{\odot}$
and different initial rotations and magnetic fields, and have calculated the evolutions.
% from the prestellar core phase to the end of the main accretion phase.
% ($i.e.,$ the end of the Class I stage).
Their simulations show that the first adiabatic core formed prior to the formation
of the protostar becomes a centrifugally-supported disk with a size of $\gtrsim$100 AU
by the end of the calculations.
% They argued that magnetic braking is not efficient enough
% to suppress the formation of the Keplerian disk, due to the effective Ohmic dissipation and the depletion of the surrounding infalling envelope.
Their results show that, however, the mass of the formed disk is considerably ($\sim$factor 2 - 100) larger
than that of the central protostar (see also Vorobyov 2009), and hence the formed disk is subject to further fragmentation. Further long-term simulations, including calculations of
fragmentations of the massive disks and subsequent evolutions, are required to
theoretically reproduce the observed Keplerian disks around the Class I protostars.

% On the other hand, Table \ref{tbl-2} demonstrates that the central protostellar masses
% dominate not only the disk masses but also the envelope masses, and thus their simulations do not reproduce
% the observationally-identified Keplerian disks.

% Simple analytical theories of star formation suggest that centrifugally supported disks
% should start out quite small (r<10 AU) and consequently with very low-mass, and
% grow with time (Terebey et al. 1984). More recent simulations of self-gravitating and viscous
% disks indicate that relatively massive disks can form at very early times (Vorobyov 2009).
% Such a scenario is supported by recent hydrodynamic simulations by Vorobyov 2009,
% which find time-averaged disk masses in Class 0 and Class I of 0.1 M for both viscous
% and self-gravitating disks.
%
% Magnetically supported disks can also be much larger (radii up to 1000 AU;
% Galli \& Shu 1993), and thus more massive at early times.
Although a Keplerian circumbinary disk is identified around L1551 NE,
L1551 NE is associated with active jets \citep{rei00,hay09} and molecular outflows \citep{mo95a},
suggesting presence of accreting circumstellar disks around the binary companions.
The central compact dust emission (Figure \ref{cont}) may indeed arise from the circumstellar disks.
% In the case of L1551 IRS 5, Lim \& Takakuwa (2006) fully
% resolved the circumstellar disks around the two main binary companions with the VLA 7-mm observations
% at an angular resolution of 0$\farcs$04 ($\sim$5 AU), and found
% the size of the circumstellar disks to be $\sim$17 AU, a few times smaller than the binary separation (= 47 AU).
% At the same time,
% high-resolution millimeter continuum imaging of L1551 IRS 5 has shown the presence of a $\sim$100 AU-scale
% circumbinary disk distinct from the circumstellar disks \citep{loo97},
% and SMA observations of L1551 IRS 5 in
% the CS (7--6) line have revealed a $\sim$400-AU scale rotating circumbinary disk \citep{tak04}.
%
% Around GG Tauri, a T-Tauri binary system, Guilloteau et al. (1999) have found a $\sim$500-AU scale
% Keplerian circumbinary ``ring''
% plus compact unresolved components which probably arise from the circumstellar disks around the
% binary companions, from their sub-arcsecond resolution millimeter continuum and $^{13}$CO (2--1)
% and HCO$^{+}$ (1--0) observations.
% Machida et al. (2008) have suggested that fragmentation of the adiabatic phase (i.e. first core)
% lead to the formation of multiple systems
%
% So far, there are no theoretical studies that can reproduce a Keplerian circumbinary disk plus accreting circumstellar disks
% in a protobinary system.
Simulations of fragmentations of a first core or a massive disk into multiple protostellar systems ($i.e.$, Machida et al. 2008)
and the further evolutions are required to reproduce a Keplerian circumbinary disk plus accreting circumstellar disks
in a protobinary system.
In the case of binary protostars, tidal effects from the two protostars can transport rotational angular momenta outwards
and create a ``gap'' in the circumbinary disk \citep{bat97,och05}.
Furthermore, the inward transportation of magnetic fields in binary systems may be quite different from that around single protostars,
because the mass accretion toward the binary protostars can occur through ``spiral'' accretion streams that connects between
the inner edge of the circumbinary disk and the circumstellar disks \citep{bat97,och05}.
The high-velocity $^{13}$CO components shown in Figure \ref{br1318} may correspond to
such accretion streams.
If the angular momentum transported outwards piles up at the inner edge of the circumbinary disk,
this disk can then assume Keplerian rotation.

% it is possible that the circumbinary disk turns to be a Keplerian disk.
% Higher-resolution observations of L1551 NE with ALMA and / or EVLA, which can separate between
% the circumstellar disks and the circumbinary disk and resolve the accretion streams connecting
% between the circumbinary disk and the circumstellar disks, are required to fully understand
% the origin of the Keplerian circumbinary disk plus the circumstellar accreting disks.

Finally, we will comment on an interesting application of Keplerian circumbinary disks.
By measuring the position of the Keplerian rotation center with respect to the binary positions,
in principle it is also possible to derive binary mass ratios, orbital radii, and orbital periods
of binary protostars.
Our $\chi^2$ model fitting of the geometrically-thin Keplerian disk
with the fixed rotational center at the position of Source A provides a reasonable fitting result.
This result may suggest that the mass of Source A is higher than that of Source B,
although our search for the minimum $\chi^2$ point of the rotational center failed to
obtain a statistically significant result (at the positions of Sources A and B the reduced $\chi^2$ values
are $\sim$1.5 and $\sim$1.8, respectively.). If we could pinpoint the rotational center,
the binary mass ratio can simply be estimated from the ratio of the distances from the rotational center to the primary and to
the secondary. This is potentially a strong method to measure mass ratios of protostellar binaries,
since previous multi-epoch ($>$ 5 yr), high-resolution ($\sim$0$\farcs$1) observations
of proper motions of protobinary systems could only deduce the reduced masses \citep{loi02,rod03,lim06}.
We can also measure the de-projected orbital radii assuming that the binary stars are in the same plane
of the circumbinary disks. From the binary masses, mass ratios, and the de-projected orbital radii
the orbital periods can also be obtained. These measurements could
put stringent constraints on theories of binary formation.

\section{Summary}

We have carried out sub-arcsecond resolution SMA observations of the protobinary system L1551 NE
in the 335 GHz continuum, $^{13}$CO (3--2), and the C$^{18}$O (3--2) emission.
The main results are summarized below:

1. The 335 GHz continuum emission in L1551 NE shows an intense central peak close to the protobinary,
and secondary peaks to the north and south of the protobinary system
located almost symmetrically with respect to the position of the protobinary.
The central component may arise from circumstellar disks around each binary companion.
The northern and southern components comprise a $\sim$600 AU scale circumbinary disk,
approximately perpendicular to the axis of the associated [Fe II] jets driven by Source A.
The mass of the circumstellar disk is estimated to be $\sim$0.02 - 0.09 $M_{\odot}$,
and that of the circumbinary disk $\sim$0.03 - 0.12 $M_{\odot}$.

2. The extent and position angle of the
$^{13}$CO (3--2) and C$^{18}$O (3--2) emission
are similar to those of the 335 GHz continuum emission,
although the locations of the emission peaks are different among the
$^{13}$CO, C$^{18}$O, and the 335 GHz emission.
Both the $^{13}$CO and C$^{18}$O emission show a systematic velocity gradient
along the major axis, which is perpendicular to the axis of the [Fe II] jets, such that their blueshifted and the redshifted
emission are located to the north and south of the protostellar binary, respectively.
% The velocity gradient in the $^{13}$CO and C$^{18}$O emission
% can be reproduced by a Keplerian rotation curve with a central stellar mass of 0.8 $M_{\odot}$,
% but not a rotation curve with the conserved angular momentum expected in the infalling gas.
In addition to this global velocity gradient, the high-velocity ($\gtrsim$3 km s$^{-1}$)
$^{13}$CO emission exhibits
compact ($\sim$100 AU) blueshifted and redshifted components to the north-west and south-east of Source A
with a different position angle from that of the circumbinary disk.
We speculate that the high-velocity $^{13}$CO components correspond to accretion streams that connects between
the inner edge of the circumbinary disk and the circumstellar disks.

3. $\chi^2$ model fitting of geometrically-thin Keplerian disks to the observed $^{13}$CO and
C$^{18}$O velocity channel maps was conducted.
The C$^{18}$O velocity channel maps are well-reproduced by a geometrically-thin Keplerian
disk model with a central stellar mass of 0.8$^{+0.6}_{-0.4}$ $M_{\odot}$,
disk position angle of 167$\degr$$^{+23\degr}_{-27\degr}$, and a disk inclination angle
of -62$\degr$$^{+25\degr}_{-17\degr}$, and a rotational center at the position of Source A.
On the other hand, the $^{13}$CO velocity channel maps cannot be fitted with any geometrically-thin
Keplerian disk model. This is probably because the $^{13}$CO emission is optically thick
and is likely affected by the contamination from the outflow and the extended envelope component.

4. Among the Class I sources with identified Keplerian disks,
L1551 NE is likely the youngest source, since L1551 NE
has the lowest bolometric temperature ($\sim$91 K), highest envelope mass ($\sim$0.39 $M_{\odot}$),
and the lowest ratio in stellar mass to envelope + disk + stellar mass ($\sim$0.65).
We suggest that L1551 NE has just passed from the ``pseudo-disk'' phase
seen around Class 0 protostars to the Keplerian-disk phase.
Our SMA observations of L1551 NE provide firm evidence that Keplerian disks around
binary protostars, Keplerian circumbinary disks, are present
as well as Keplerian disks around single protostars.

5. Current theoretical studies cannot reproduce observationally-identified Keplerian disks around single or multiple
protostars. In the case of binary protostars, tidal effects from two protostars can transport rotational angular momenta
outward, and pile up the angular momenta at the inner edge of the circumbinary disk effectively.
Furthermore, the inward transportation of magnetic fields in binary systems may be quite different from that around single protostars,
because the mass accretion toward the binary protostars can occur through ``spiral'' accretion
streams from the inner edge of the circumbinary disk to the circumstellar disks.
These mechanisms could create a Keplerian circumbinary disk around binary protostars.
Higher-resolution observations of L1551 NE with ALMA and / or EVLA, which can separate between
the circumstellar disks and the circumbinary disk and resolve the accretion streams connecting
between the circumbinary disk and the circumstellar disks, are required to fully understand
the origin of the Keplerian circumbinary disk plus the circumstellar disks.

\acknowledgments
We are grateful to P. T. P. Ho and N. Ohashi for their fruitful discussions. We would like to thank all the SMA staff supporting this 
work. S.T. acknowledges a grant from the National Science
Council of Taiwan (NSC 99-2112-M-001-009-MY3) in support of this work.

\clearpage
%%%%%%%%%%%
% Figures %
%%%%%%%%%%%

% \begin{figure}
% \epsscale{1.0}
% \plotone{KepfitR.eps}
% \caption{Same as Figure \ref{kepb} but for the redshifted velocities.
% \label{kepr}}
% \end{figure}

%%%%%%%%%%%
% Table 1 %
%%%%%%%%%%% 
\clearpage
\begin{deluxetable}{lcc}
\tabletypesize{\scriptsize}
\tablecaption{Parameters for the SMA Observations of L1551 NE \label{tbl-1}}
\tablewidth{0pt}
% \tablewidth{250pt} % use this in emulateapj
\tablehead{\colhead{Parameter} & \multicolumn{2}{c}{Value}\\
\cline{2-3}
\colhead{} & \colhead{2010 January 15} & \colhead{2010 January 18} }
\startdata
Number of Antennas &8 & 7 \\
%Field Center &  &  \\
Right ascension (J2000.0)
   & \multicolumn{2}{c}{04$^{\rm h}$ 31$^{\rm m}$ 44$^{\rm s}$.47}\\
Declination (J2000.0) 
   & \multicolumn{2}{c}{18$^{\circ}$ 08$\arcmin$ 32\farcs2}\\
Primary Beam HPBW& \multicolumn{2}{c}{$\sim$37$\arcsec$}\\
Synthesized Beam HPBW (Line)&
  \multicolumn{2}{c}{0\farcs95 $\times$ 0\farcs66 (P.A. = -88$\degr$)}\\
Synthesized Beam HPBW (Continuum)&
  \multicolumn{2}{c}{0\farcs80 $\times$ 0\farcs54 (P.A. = -87$\degr$)}\\
Baseline Coverage & \multicolumn{2}{c}{29 - 249 (k$\lambda$)}\\
Conversion Factor (C$^{18}$O) & \multicolumn{2}{c}{1 (Jy beam$^{-1}$) = 17.9 (K)}\\
Frequency Resolution & \multicolumn{2}{c}{406.25 kHz $\sim$0.37 km s$^{-1}$}\\
Bandwidth &  \multicolumn{2}{c}{7.74 GHz}\\
Flux Calibrator &Titan &Callisto\\
Gain Calibrator & \multicolumn{2}{c}{0423-013, 3c120}\\
Flux (0423-013 Upper) &3.46 Jy &3.62 Jy\\
Flux (0423-013 Lower) &3.53 Jy &4.42 Jy\\
Flux (3c120 Upper)       &0.80 Jy &0.92 Jy\\
Flux (3c120 Lower)       &0.83 Jy &1.08 Jy\\
Passband Calibrator    &\multicolumn{2}{c}{3c273}  \\
System Temperature (DSB) &$\sim$300 - 700 K &$\sim$400 - 1000 K\\
rms noise level (Continuum)& \multicolumn{2}{c}{3 mJy beam$^{-1}$}\\
rms noise level (C$^{18}$O)& \multicolumn{2}{c}{0.117 Jy beam$^{-1}$}\\
rms noise level ($^{13}$CO)& \multicolumn{2}{c}{0.107 Jy beam$^{-1}$}\\
\enddata
\end{deluxetable}

\clearpage
\begin{deluxetable}{lcccccccc} \tablecaption{Protostellar Sources with Keplerian Disks\label{tbl-2}}
\tabletypesize{\scriptsize}
\rotate
\tablewidth{0pt}
\tablehead{\colhead{Source} &\colhead{$L_{bol}$} &\colhead{$T_{bol}$} &\colhead{$R_{kep}$$\tablenotemark{a}$} &\colhead{$M_{star}$$\tablenotemark{b}$}
&\colhead{$M_{disk}$$\tablenotemark{c}$} &\colhead{$M_{env}$$\tablenotemark{d}$}
&\colhead{$\frac{M_{star}}{M_{star}+M_{disk}+M_{env}}$} &\colhead{references$\tablenotemark{e}$}\\
\colhead{} &\colhead{($L_{\odot}$)} &\colhead{(K)} &\colhead{(AU)} &\colhead{($M_{\odot}$)} &\colhead{($M_{\odot}$)} &\colhead{($M_{\odot}$)} &\colhead{($\%$)} &\colhead{}}
\startdata
L1551 NE  &4.2   &91   &300 &0.8   &0.026$\tablenotemark{f}$            &0.39   &65 &1,2,3\\
L1489-IRS &3.7   &238 &200 &1.35 &0.004            &0.093 &93 &4,5\\
IRS 43      &6.0   &310 &140 &1.0   &0.0081          &0.026 &97  &5  \\
IRS 63      &0.79 &351 &100 &0.37 &0.055            &0.022 &83 &5,6  \\
Elias 29    &13.6 &391 &200 &2.5   &$\leq$0.007  &0.025 &$\geq$99  &5,6  \\
\enddata

\tablenotetext{a}{Outer radius of the Keplerian disk.}
\tablenotetext{b}{Central protostellar mass derived from the Keplerian-Disk model fitting.}
\tablenotetext{c}{Mass of the Keplerian disk measured from the SMA continuum flux.}
\tablenotetext{d}{Mass of the protostellar envelope estimated from the Single-dish submillimeter continuum flux.}
\tablenotetext{e}{References: (1) This work; (2) Froebrich 2005; (3) Moriarty-Schieven et al. 2006;
(4) Brinch et al. 2007; (5) J$\o$rgensen et al. 2009; (6) Lommen et al. 2008}
\tablenotetext{f}{Excluding the mass of the central compact component that may arise from the circumstellar disks
around the binary companions.}
\end{deluxetable}

\end{document}